\documentclass{elsart}
\usepackage{epsfig,rotating}

\def \slash#1{\centeron{$#1$}{$/$}}
\def\centeron#1#2{{\setbox0=\hbox{#1}\setbox1=\hbox{#2}\ifdim
   \wd1>\wd0\kern.5\wd1\kern-.5\wd0\fi
   \copy0\kern-.5\wd0\kern-.5\wd1\copy1\ifdim\wd0>\wd1
   \kern.5\wd0\kern-.5\wd1\fi}}
\newcommand{\lsim}   {\mathrel{\mathop{\kern 0pt \rlap
  {\raise.2ex\hbox{$<$}}}
  \lower.9ex\hbox{\kern-.190em $\sim$}}}
\newcommand{\gsim}   {\mathrel{\mathop{\kern 0pt \rlap
  {\raise.2ex\hbox{$>$}}}
  \lower.9ex\hbox{\kern-.190em $\sim$}}}
\def\be{\begin{equation}}
\def\ee{\end{equation}}
\def\ba{\begin{eqnarray}}
\def\ea{\end{eqnarray}}
\def\i{{\rm i}}
\def\Sp{{\rm Tr}}
\def\sp{{\rm tr}}
\def\E0{E_{\rm vac}}
\def\eps{\varepsilon}

\begin{document}
\begin{frontmatter}

\title{Neutrino self-energy  and pair creation in neutron stars}
\author{M. Kachelrie{\ss}}

\address{INFN, Laboratori Nazionali del Gran Sasso,
             I--67010 Assergi (AQ), Italy}

\begin{abstract}
Massless neutrino exchange leads to a new long-range force between matter.
Recently, it was claimed both that the potential energy due to this
interaction i) dominates the total energy of neutron stars and 
ii) that it is zero. We recalculate the energy of a neutrino
propagating in a classical, uniform background of 
neutrons and find a negligible, but non-zero contribution to the total
energy of neutron stars. 
We estimate the neutrino pair creation rate of a neutron star caused 
by a density gradient of the background neutrons 
but found it too small to be observable.
\end{abstract}
\end{frontmatter}

\section{Introduction} 
Recently Fischbach calculated the energy difference $\Delta E$
between a neutrino immersed in a neutron star and in vacuum \cite{fi96}. 
He used perturbation theory to derive the potential energy $W^{(k)}$ 
of $N$ neutrons due to the exchange of massless neutrinos, $\Delta
E=\sum_k W^{(k)}$.
The ratio of the contributions  to $\Delta E$ from $k$ and $k+2$ body
interactions found by him equals for $k\ll N$
\be
 \left| \frac{W^{(k+2)}}{W^{(k)}} \right| 
 \sim \frac{1}{(k+2)(k+1)} 
      \left( \frac{G_F N}{R^2} \right)^2 \,,
\ee
where $G_F$ is the Fermi constant and $R$ the radius of the star.
For a typical neutron star, $G_F N /R^2 = {\mathcal O}(10^{13})$ and
multi-body 
effects become dominating. In particular, $|\Delta E|$ exceeds the mass
energy of the neutron star---an obvious contradiction to the
observation of neutron stars. The resolution of 
this paradox proposed in Ref. \cite{fi96} is to consider a massive neutrino. 
Then the neutrino can interact  only with neutrons within its Yukawa
radius $1/m_\nu$, and, if $m_\nu$ is sufficiently large, the dangerous
many-body effects are exponentially damped. Hereby, a lower bound for
the electron neutrino mass, $m_{\nu_e} \gsim 0.4$~eV, was derived \cite{fi96}.

The method and results of Ref.~\cite{fi96} provoked some criticism. 
Smirnov and Vissani \cite{sm96} argued that a neutrino sea inside the
neutron star \cite{lo90} reduces the potential energy $W^{(k)}$
because of Pauli blocking. They stressed also that the behaviour of
the potential energy $W^{(k)}$,
\be
 W^{(2k)}= (-1)^{k} |\tilde W^{(2k)} |  \,,
\ee
is unacceptable. In fact, both the increasing of $|W^{(k)}|$ and its
oscillatory behaviour with $2k$ are clear signs for the
breakdown of perturbation theory.
Abada {\it et al.\/} \cite{ab96} recalculated the energy difference
$\Delta E$ taking into account non-perturbatively the interaction of the
neutrino with a uniform neutron background. They obtained
$\Delta E=0$ if no neutrino sea is presented. This result would imply
that neutrinos do not interact at all with the neutron
background. In view of the results of
Ref. \cite{fi96,ab96} and the possible implications for neutrino
physics, we feel it appropriate to recalculate once again $\Delta E$.
Additionally, we estimate the neutrino pair creation rate of a neutron
star caused by a density gradient of the background neutrons.

In our calculations we always assume
that the neutron density $n_N({\bf x})$ respectively its gradient 
${\bf \nabla} n_N({\bf x})$ can be approximated locally by a constant
value. This assumption looks unproblematic because of the
macroscopic size of the neutron star, neglects however
possible long-range effects due to massless neutrinos.

\section{Neutrino energy density without neutrino sea}
The energy $E$ of a neutrino interacting with a classical background
current $J^\mu$ of neutrons is given by \cite{sc54,gr85}
\be
 E = \langle P^0 \rangle
   = \i \int\d^3 x \:\langle 0 | \psi^\dagger (x) \partial_t \psi(x)|0 
                     \rangle_J \,.
\label{E}
\ee
Here, $\psi$ is the field operator in the Furry picture obtained after
second quantizing the solutions of the Dirac equation
%
\be
 \left[ \gamma_\mu\partial^\mu +\i m
  - \sqrt{2} \i G_F a_N \gamma_\mu J^\mu P_L \right] \psi(x)=0 \,,
\label{DG}
\ee
where $a_N=-0.5$ and $P_L=(1-\gamma^5)/2$ projects out
the right-handed component of the neutrinos.

Since we are interested in the infrared regime, we can assume in the
following $J^\mu$ as static, $J^\mu=(n_N,{\bf 0})$. 
Moreover, the neutron number
density $n_N$ is also nearly homogenous. Therefore, the solutions of
Eq.~(\ref{DG} ) can still be characterized by the four-momentum
$p^\mu=(E,{\bf p})$ of the neutrino. Then, as it is well known, the only 
change compared to the vacuum case is the modification of the dispersion
relation of the neutrino \cite{wo78,vz},
\be
 E =  (m^2+|{\bf p}|^2)^{1/2} \pm V 
\label{disp}
\ee
\be
 V = - \frac{G_F}{\sqrt{2}}\: n_N \,.
\ee
Here, the upper sign in Eq.~(\ref{disp}) corresponds to neutrinos and
the lower sign to antineutrinos. 

Equation (\ref{E}) can be rewritten as
\be
 E = \int\d^3 x\:  \partial_t \sp \left\{ \gamma^0 S_F(x,x') 
                                    \right\}_{x'\searrow x} \,,
\label{S}
\ee
where $S_F(x,x')$ is the Greens function of Eq.~(\ref{DG}). 
Now we evaluate the energy difference 
\be
 \Delta E =  \int\d^3x \int\frac{\d p^4}{(2\pi)^4} 
             \: (-\i p^0) e^{-\i p (x-x')} \:
             \sp\left\{ \gamma^0 \left[ S_F(p) - S_F^{(0)} (p) \right] 
                                     \right\}_{x'\searrow x}
\ee
between a neutrino propagating in a neutron background \cite{prop},
\be
 S_F (p) = \frac{1}{(p^0 - V)\gamma^0 - {\bf p} \cdot {\bf \gamma}-m+\i\eps}
           \: P_L \,,
\ee
and a neutrino propagating in the vacuum,
\be
 S_F^{(0)} (p) = \frac{1}{\slash p-m+\i\eps} \: P_L \,.
\ee
Performing first the ${\bf x}$ and then the ${\bf p}$ integral results in
\be
 \Delta E =  -\frac{\i}{2\pi} \int\d p^0 \: p^0 e^{-\i p^0 (t-t')} \:
   \sp \left\{ \gamma^0 \left[ S_F(p^0,{\bf 0}) 
                             - S_F^{(0)} (p^0,{\bf 0}) \right] 
                                     \right\}_{t'\searrow t}  \,.
\label{de}
\ee
%
Although each of the two contributions to the energy difference
$\Delta E$ is UV-divergent, a finite final result for $\Delta E$ 
can be obtained combining the two fractions before integrating.
Performing first the subtraction regularizes the integral, because 
then its leading term in $p^0$ vanishes, 
\be
 \Delta E =   -\frac{\i V}{\pi} \int \d p^0 \: p^0 e^{-\i p^0 (t-t')} \:
             { \frac{(p^0)^2 - p^0 V +m^2}
                    { [(p^0-V)^2-m^2+\i\eps] [(p^0)^2-m^2+\i\eps] }
             }_{t'\searrow t} .
\ee
%
The final integral can be done with the help of the residue
theorem (cf. Fig.~1) and gives independently of the neutrino mass $m$ 
\be
 \Delta E = V \approx -19~{\rm eV} \:\frac{n_N}{0.3~{\rm fm}^{-3}} .
\ee
This result could be expected in virtue of Eq.~(\ref{disp}). 
Since $\Delta E$ is the energy difference of one neutrino, we obtain
the total contribution $\Delta M$ of neutrinos to the self-energy of a
neutron star with radius $R\approx 10$~km as 
\be
 \Delta M = \frac{4\pi}{3}R^3 n_\nu \Delta E   \approx -42\:{\rm kg}
 \; \frac{n_\nu}{3 \times 10^{-22}\:{\rm fm}^{-3}} .
\ee
Obviously, $\Delta M$ is negligible for a realistic value (cf. next
sections) of the neutrino density $n_\nu$.

Finally, we want to comment on the possible source of error in the
results of Ref. \cite{fi96,ab96}. In Ref. \cite{fi96}, the 8th order term of 
perturbation theory $W^{(8)}$ was used to estimate $\Delta E$. Since 
$\sum_k W^{(k)}$ is divergent and alternating, any 
result obtained in a finite order of perturbation theory is meaningless.   
By contrast, the authors of Ref. \cite{ab96} used the same non-perturbative 
method as we did. However, they made -- as apparent from their Eq.~(8) -- 
the limit $x' \searrow x$ {\em before\/} integrating over $x$. Then,
they argued that after regularization their $\d^4p$ integral
vanishes due to the antisymmetry of the integrand. However, this is
not true because of the presence of the factor $p^0 e^{-ip^0(t-t')}$.

\section{Neutrino energy density with neutrino sea}
Loeb proposed first that neutrinos%
\footnote{In Ref. \cite{ki97}, it is argued that 
only Dirac neutrinos are trapped while Majorana
neutrinos can leave freely the medium.}
with energies below $\sim
50$~eV are bounded inside neutron stars, while antineutrinos are
repelled \cite{lo90}. 
Qualitatively, this result follows directly from Eq. (\ref{disp}) 
considering a neutron star as potential wall with depth $V$ and radius
$R$, and simply
assuming that all the levels with energy $[0,V]$ are occupied. 
However, one should note that neutrino states with energy $V<E<-m_\nu$ do
not have an exponentially damped wave function for $r>R$, and are
resonances instead of true bound states.

To account for a possible neutrino sea  
inside an neutron star, we can apply either the
imaginary or the real-time formalism of finite-temperature field
theory. We use the latter since it is easier in this formalism to
separate the medium from the vacuum effects. Then
the thermal neutrino propagator $S_F(p,\beta)$ is given by
\ba
 S_F (p,\beta) & = & \left[ (p^0 -V)\gamma^0 - {\bf p \cdot \gamma}
                     \right]
 \Big[  \frac{1}{(p^0 -V)^2 - |{\bf p}|^2}
\nonumber\\ & + &
  2\pi\i \delta( (p^0 -V)^2 - |{\bf p}|^2 ) f(p^0)  \Big] \: P_L \,,
\label{thermo}
\ea
where $\beta=1/T$ is the inverse temperature and $f(p^0)$ is the
distribution function of the neutrinos. To simplify the
notation, we have set $m=0$. In the case of thermal
equilibrium, $f(p^0)$ is the Fermi-Dirac distribution function
\be
 f_D (p^0) = f_{D,\nu} (p^0) + f_{D,\bar\nu} (-p^0)
           = \frac{\theta(p^0-V)}{e^{\beta(p^0-\mu)}+1} 
           + \frac{\theta(-p^0-V)}{e^{-\beta(p^0-\mu)}+1} \,.
\label{f_D}
\ee
Note that the edge between particles and antiparticles is shifted by $V$.
Inserting the thermal propagator into Eq.~(\ref{S}), performing first
the ${\bf x}$ integration restricted to the volume ${\mathcal V}$ of the
neutron star and then the $p_0$ integral results in
\ba
 \Delta E_\beta / {\mathcal V} 
          &  = & \int \frac{\d^3 p}{(2\pi)^3} \:
                 \left[ (|{\bf p}|+V) f_{\nu} (|{\bf p}|+V) 
                      + ({-|\bf p}|+V) f_{\bar\nu} (|{\bf p}|-V) 
                 \right] 
\\
          &  = & \int \frac{\d^3 p}{(2\pi)^3} \:
                 \left[ E_\nu f_{\nu} (E_\nu) 
                      - E_{\bar\nu} f_{\bar\nu} (E_{\bar\nu}) 
                 \right] 
\\  & = &
     \langle E_\nu \rangle_\beta -  \langle E_{\bar\nu} \rangle_\beta .
\label{E_beta}
\ea
In the last step, we denoted the thermal average with the
(anti-)particle distribution functions by $\langle\ldots\rangle_\beta$.
Also in this case, the result obtained has a very plausible form.

To obtain a numerical estimate for $\Delta E_\beta$, we follow
Ref.~\cite{lo90} and assume a
degenerated neutrino sea with $n_\nu\sim 3\times 10^{-22}$~fm$^{-3}$ 
and Fermi momentum $p_F\sim 50$~eV. Furthermore we set
$n_{\bar\nu}=0$ and obtain
\be
 \Delta E_\beta / {\mathcal V} \sim \left( V + \frac{3}{4} p_F \right) n_\nu
 \sim 5.5\times 10^{-21} \, \frac{\rm eV}{{\rm fm}^3} \:
      \frac{n_\nu}{3\times 10^{-22}~{\rm fm}^{-3}} \,. 
\ee
For a neutron star, the thermal contribution
$\Delta E_\beta$ is dominated by the
contribution of the kinetic energy of the neutrinos and therefore positive.
However, since the number density of neutrinos is much smaller than the  
number density of neutrons, $n_\nu/n_N\sim 10^{-21}$, the thermal
contribution $\Delta E_\beta$ is irrelevant compared to $\Delta E$.
Finally, we want to remind that -- as mentioned in the introduction --
the results obtained in section 2 and
3 are only valid for an uniform neutron background, {\it i.e.\/} in
the limit of an infinite neutron star.

\section{Spontaneous neutrino pair creation}
Let us consider in more detail the analogy between a neutron star and
a potential wall. We keep now again the neutrino mass $m_\nu$ finite and
consider Dirac neutrinos. 
The limit for the electron neutrino mass is $m_{\nu_e} \lsim
5$~eV. Hence the potential\footnote{Since the absolute value of the
potential $V$ has no physical meaning, a more correct statement is that
the potential difference $\Delta V=V(r<R)-V(r>R)$ is overcritical.}
$V\sim -50$~eV is overcritical, $V<-m_{\nu_e}$, and therefore able to produce
$\nu_e\bar\nu_e$-pairs at its interface. 
The quantity characterizing this process, $m_\nu^2/|{\bf \nabla} V|$, 
is  not well-defined in the simple picture of the potential wall.
We assume instead that $n_N({\bf x})$ can be approximated by 
$n_N({\bf x})=n_N({\bf x}_0) + ({\bf x}-{\bf x}_0) \nabla n_N ({\bf x})$.
Then we can treat locally the ``electric'' field 
${\bf E}=-\nabla V-\partial_t {\bf A}$, where $A=(V,{\bf 0})$, as
uniform and derive the spontaneous neutrino pair creation rate due to a
density gradient of the neutron background. 

The action $S(A)$ describing the vacuum with a potential $A$ is given
by \cite{sc49,it80}
\be
 \ln S(A) = \Sp\ln\left\{  S_F^{-1}  S_F^{(0)} \right\} \,.
\label{S(A)}
\ee
Here, $\Sp$ means the trace over Dirac indices and integration over
the continuous variable $x$. Furthermore, $S_F$ denotes the operator
with matrix element $\langle x| S_F | x' \rangle = S_F(x,x')$.
Adding the transposed version of its RHS to Eq.~(\ref{S(A)}),
inserting $CC^{-1}={\bf 1}$ and using
$C\gamma^5 C^{-1}=\gamma^{5,t}$, $C\gamma_\mu C^{-1}=-\gamma_\mu^{t}$,
we obtain
\ba
 2\ln S(A) & = & \Sp\ln\bigg\{ 
                   P_L ( \slash P - \slash A +m_\nu -\i\eps )
                   ( \slash P - \slash A -m_\nu +\i\eps ) P_L
\\ &\times&
	           \frac{1}{\slash P -m_\nu +\i\eps} P_L
                   \frac{1}{\slash P +m_\nu -\i\eps} 
                     \bigg\}  
\nonumber\\
 & = & 
     \frac{1}{2} \Sp\ln\left\{ \left[ 
     ( P - A )^2 -m^2_\nu +\i\eps + \frac{1}{2} \sigma_{\mu\nu}F^{\mu\nu} 
     \right] \:  \frac{1}{P^2 -m^2_\nu +\i\eps} \right\}  
\ea
This is $1/2$ of the corresponding result 
for the pair production of fermions by an uniform electric field. Therefore,
we can borrow the final QED result \cite{it80} and obtain for the pair
creation probability per unit time and volume
\be
 w = \frac{E^2}{8\pi^3} \sum_{n=1}^\infty \frac{1}{n^2}
                         \exp\left( -\frac{n\pi m_\nu^2}{|E|} \right) \,.
\ee
To obtain an order-of-magnitude estimate for $w$, we choose a density
profile that models roughly  a neutron star with a soft Reidl 
equation-of-state \cite{sh83}: we assume as radius of the star
$R=10$~km, a homogenous core and a crust with thickness
$L=2$~km, in which the density decreases linearly to zero,
\be
 n_N(r)= \left\{ \begin{array}{ll} n_0 & \quad\textrm{ for} \quad r<R-L \\
                                n_0 (R-r)/L  & \quad\textrm{ for} \quad 
                                                   R-L < r < R \,.
               \end{array} \right. 
\ee 
Then, in a volume ${\mathcal V} \sim 2\cdot 10^{18}$~cm$^3$ of the
neutron star exists a field $|E|= V/L\sim 5\cdot 10^{-9}$~eV$^2$ in
radial direction.
If $m_\nu^2 \gsim |E|/(n\pi)$, the creation of a $\nu\bar\nu$-pair by
$n$ field quanta is exponentially suppressed. But even if 
$m_\nu \ll (|E|/\pi)^{1/2} = 4\cdot 10^{-5}$~eV, the total luminosity
${\mathcal L}$ is only
\be
 {\mathcal L} = 2m_\nu w {\mathcal V} \sim
                2.1 \cdot 10^{11} {\rm erg/s} \left(
                \frac{m_\nu}{10^{-6}{\rm eV}} \right)
\ee
and has therefore practically no influence on the energy budget of the
star. However, $\nu_e
\bar\nu_e$-pair creation by neutron stars is probably the only case in
the present Universe that massive particles are spontaneously created
and therefore it is interesting in its own.

\section{Summary}
We have recalculated the energy of a neutrino in the uniform background of
classical neutrons, both without and with a neutrino sea.
In the first case, we found that the energy is changed by the small
amount $V\sim -20$~eV, in agreement with the well-known result of
Wolfenstein \cite{wo78,vz}. Moreover, the influence of
a possible neutrino sea inside the neutron star
has an even smaller effect on the neutrino energy.
Therefore, the contribution of neutrinos to the total energy of a
neutron star seems to be negligible. However, to settle definitely
the question if many-body effects become important in neutron stars
it is necessary to calculate $\Delta E$ not only non-pertubatively but to
take into account also the finite geometry of the star \cite{fu}.
We have estimated the energy-loss of an neutron star due to spontaneous
$\nu\bar\nu$-pair production but found it too small to be observable.

\begin{ack}
I am grateful to E. Fischbach and K. Kiers for helpful remarks, and
especially to V. Semikoz, who pointed out an error in an early version
of the manuscript.
This work was supported by a Feodor-Lynen scholarship of the Alexander von 
Humboldt-Stiftung.

After submission of this work, a preprint of As. Abada, O. P{\`e}ne
and J. Rodrigues-Quintero (hep-ph/9712266) following the line of
arguments of Ref.~\cite{ab96} appeared. Their main conclusion is
that the border of a finite neutron star (modelled as a square wall)
automatically generates the neutrino sea inside the star.
\end{ack}


\newpage

\begin{center}
\begin{figure}
\setlength{\unitlength}{1cm}
\begin{center}
\begin{picture}(12,0)
 \epsfig{file={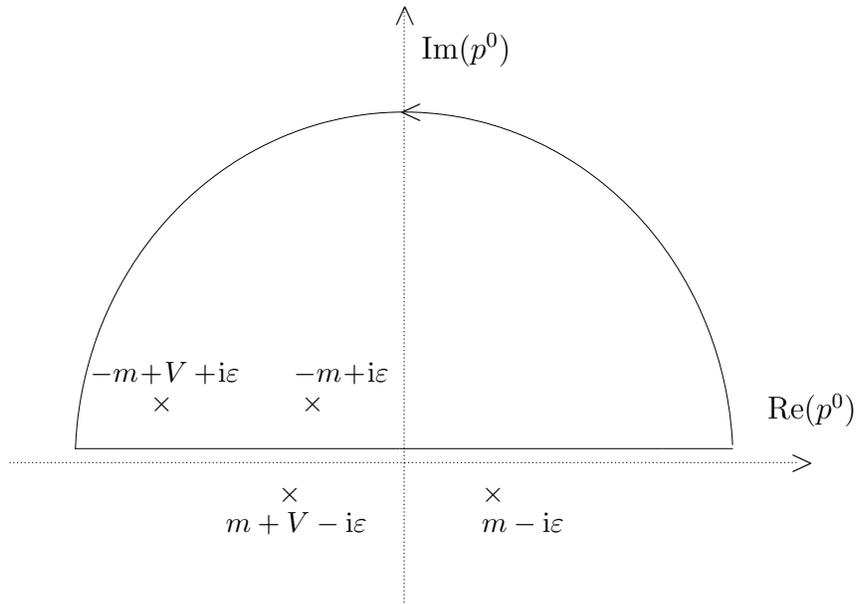}, height=12cm, angle=270}
 \put (-5.4,-0.8){$\mathrm{Im}(p^0)$} 
 \put (-5.74,-0.4){\begin{sideways}
                  $>$
                  \end{sideways}} 
 \put (-0.8,-5.6){$\mathrm{Re}(p^0)$} 
 \put (-0.5,-6.3){$>$} 
 \put (-5.7,-1.62){$<$} 
 \put (-7.1,-5.1){{\footnotesize $-m\!+\!\i\eps$}} 
 \put (-7.0,-5.5){{\footnotesize $\times$}} 
 \put (-4.6,-7.1){{\footnotesize $m-\i\eps$}} 
 \put (-4.6,-6.7){{\footnotesize $\times$}} 
 \put (-9.8,-5.1){{\footnotesize $-m\!+\!V+\!\i\eps$}} 
 \put (-9.0,-5.5){{\footnotesize $\times$}} 
 \put (-8.0,-7.1){{\footnotesize $m+V-\i\eps$}}
 \put (-7.3,-6.7){{\footnotesize $\times$}}  
\end{picture}
\end{center}
\vskip8.0cm
 \caption{Residues and contour of integration for the evaluation of
 $\Delta E$, Eq.~(12).}
\end{figure}
\end{center}

\end{document}